\documentclass[%
nofootinbib,
 amsmath,amssymb,
 aps,
 prd,
twocolumn, superscriptaddress
]{revtex4-1}
\usepackage[normalem]{ulem}
\usepackage{graphicx}
\usepackage{dcolumn}
\usepackage{bm}
\usepackage{hyperref}
\usepackage{footnote}
\usepackage{float}
\usepackage{xcolor}
\usepackage{physics}
\usepackage{scalerel}
\usepackage{enumitem}

\begin{document}

\title{Semi-analytic bounds on axion-like-particle supernovae emission}

\author{Ana Luisa {\sc Foguel}} \email{afoguel@usp.br}
 \affiliation{
 Instituto de Física,
Universidade de São Paulo, 05508-090 São Paulo, SP, Brazil 
}

\author{Eduardo S. {\sc Fraga}} \email{fraga@if.ufrj.br}
\affiliation{
 Instituto de F\'\i sica, Universidade Federal do Rio de Janeiro,\\
 CEP 21941-909 Rio de Janeiro, RJ, Brazil 
}

\def\be{\begin{equation}}
\def\ee{\end{equation}}
\newcommand{\ba}{\begin{eqnarray}}
\newcommand{\ea}{\end{eqnarray}}

\def \ie{{\it i.e. }}
\def \eg{{\it e.g.}}
\def \etal{{\it et al.}}

\def \GeV{{\, \mathrm{GeV}}}
\def \MeV{{\, \mathrm{MeV}}}

\def\ALF#1{{\color{red}[#1]}} 
\def\EF#1{{\color{blue} [#1]}} 

\definecolor{myblue}{HTML}{0077b6}
\definecolor{myred}{HTML}{db5a53}
\definecolor{mygreen}{HTML}{2a9d79}
\def\myr#1{{ \textcolor{myred}{#1 } }}
\def\myb#1{{ \textcolor{myblue}{#1 } }}
\def\myg#1{{ \textcolor{mygreen}{#1 } }}

\begin{abstract} 
Core-collapse supernovae provide natural laboratories for the production of new light particles. In particular, axion-like particles (ALPs) can be constrained via SN~1987A cooling arguments. However, significant astrophysical and nuclear uncertainties imply that such bounds may vary strongly depending on modeling choices, even when expensive simulations are employed. In this context, semi-analytic methods offer a simple and fast alternative for deriving new-physics constraints. Building on a previous semi-analytic framework, in which proto-neutron star (PNS) observables are expressed in terms of six global PNS parameters, we include a finite ALP mass in the calculation and derive bounds in the axion-nucleon coupling versus mass plane. The obtained bounds are in good agreement with previous results from numerical simulations, demonstrating the robustness of the method. We also illustrate the sensitivity of the bounds to different PNS parameter calibrations, nuclear effects and cooling exclusion criteria.
\end{abstract}

\maketitle

\section{Introduction}

Physics beyond the Standard Model (SM) seems to be necessary to explain a number of unsolved fundamental questions, such as the lack of a suitable dark matter candidate, the origin of neutrino masses, and the baryon asymmetry of the Universe. Although initially new physics searches focused mostly on the heavy-mass regime, the absence of any hints of heavy dark sectors has shifted attention toward new light particles, feebly coupled to the SM. In this context, astrophysical environments serve as powerful probes of new light sectors, complementary to laboratory experiments, especially in the regime of very small couplings that are difficult to reach in colliders \cite{Raffelt:1996wa}.

Among different astrophysical environments, core-collapse supernovae represent one of the most energetic transient events in the universe, creating extreme conditions favorable for the production of light new physics. Once produced, such feebly-interacting particles can freely stream out of the stellar core, acting as an additional energy-loss channel and leaving observable imprints on the neutrino signal. The detection of neutrinos from SN~1987A~\cite{Kamiokande-II:1987idp,Bionta:1987qt} provides the only observational data available to date to constrain such effects, and has been extensively used to place bounds on a variety of new physics scenarios.

In particular, among the possible SM extensions, axion-like particles (ALPs) are especially well-motivated candidates, since they arise as pseudo-Nambu-Goldstone bosons of the spontaneous breaking of global symmetries. This makes them a common prediction of many different model-building frameworks addressing distinct open problems, such as the strong CP problem (with the QCD axion~\cite{Peccei:1977hh,Weinberg:1977ma,Wilczek:1977pj}), the origin of neutrino masses (Majorons~\cite{Chikashige:1980ui,Gelmini:1980re}), the baryon asymmetry of the Universe (via axiogenesis~\cite{Co:2019wyp} or the Bajoron~\cite{Bittar:2024nrn}, for instance), the flavor hierarchy problem (familons~\cite{Wilczek:1982rv}), and many others. Furthermore, ALPs can also play a role in the dark matter quest, either as candidates themselves~\cite{PhysRevLett.51.1415,DiLuzio:2020wdo,OHare:2024nmr} or as portal particles connecting the SM to the dark sector~\cite{Bharucha:2022lty,Armando:2023zwz,Beenakker:2025mhf}. Recent developments in axion physics include, for instance, state-of-the-art non-perturbative determination of the axion-photon coupling from lattice QCD~\cite{Brandt:2022jfk,Brandt:2026wir}.

In the literature, constraints from supernova cooling have been derived in the ALP coupling versus mass parameter space, considering different SM couplings, such as to nucleons~\cite{Turner:1987by,Carenza:2019pxu,Lella:2022uwi,Lella:2023bfb} and to photons~\cite{Payez:2014xsa,Lee:2018lcj,Lucente:2020whw}. However, many of these bounds have been obtained using computationally expensive simulations, which are time-consuming and difficult to systematically explore over a wide range of ALP parameters and modeling assumptions. Besides, even when such simulations are employed, results can vary significantly due to uncertainties in the nuclear microphysics and astrophysical inputs, such as the progenitor mass, the equation of state, or the choice of SN simulation. In this context, semi-analytic methods represent a fast and flexible alternative, allowing one to estimate the impact of ALP emission on the energy loss and to systematically explore the dependence of the bounds on some of the underlying assumptions, without the need for dedicated numerical simulations.

In this work we employ a semi-analytic method to derive constraints in the axion-nucleon coupling versus mass parameter space, $(g_{aNN},m_a)$, based on the framework developed in Refs.~\cite{Foguel:2022fef,Suwa:2020nee}. In this framework, the relevant PNS observables, such as the temperature, neutrino luminosity, and total emitted energy, are expressed in terms of six global parameters: the PNS mass $M_{\rm PNS}$, radius $R_{\rm PNS}$, density correction factor $g$, opacity boosting factor $\beta$, total neutrino-emitted energy $E_{\rm tot}$, and proton fraction $Y_p$, together with an entropy function that controls the time evolution. 

The axion luminosity from nucleon-nucleon Bremsstrahlung, expressed in terms of the same PNS quantities and the ALP parameters (extended here to include a finite ALP mass) is incorporated into the energy-loss differential equation alongside the neutrino luminosity. Solving this equation yields the entropy evolution, from which the luminosities can be computed semi-analytically, allowing us to derive bounds in the $(g_{aNN}, m_a)$ parameter space by imposing supernova cooling criteria. The obtained bounds are found to be in good agreement with previous results from full numerical simulations~\cite{Lella:2022uwi,Lella:2023bfb}, demonstrating the robustness and utility of the semi-analytic approach.

The paper is organized as follows. In Sec.~\ref{sec:method} we describe the methodology, which is divided into three parts: a review of the semi-analytic setup and the relevant formulas (Sec.~\ref{sec:analytic}), the procedure to incorporate a finite axion mass into the computation (Sec.~\ref{sec:Maxion}), and the calibration of the global PNS parameters against core-collapse supernova simulations (Sec.~\ref{sec:SNsim}). The resulting bounds in the $(g_{aNN}, m_a)$ plane are presented and compared with previous results from the literature in Sec.~\ref{sec:results}. We conclude with our final 
remarks in Sec.~\ref{sec:outlook}.

\section{Methodology} \label{sec:method}

To derive bounds in the axion-nucleon coupling versus axion mass parameter space, we proceed in three steps. After a brief review of the semi-analytic setup developed in Ref.~\cite{Foguel:2022fef}, we first incorporate a finite axion mass into the supernova axion luminosity calculation. Next, using time-dependent outputs from state-of-the-art core-collapse supernova simulations we calibrate the six global parameters of our semi-analytic PNS model by fitting to the simulated neutrino luminosities. 
With the calibrated parameters that best reproduce the numerical solutions, we compute the axion emissivity and apply supernova-cooling criteria to obtain exclusions in the  $(g_{aNN}, m_a)$ plane, which we compare with previous bounds in the literature.

\subsection{Review of analytic setup} \label{sec:analytic}

The semi-analytic setup employed in this work follows the one presented in more detail in Ref.~\cite{Foguel:2022fef}, which in turn is based on the framework developed in Ref.~\cite{Suwa:2020nee}. The key idea behind this approach for axion-like particle emission in supernovae is to express all relevant ALP and neutrino observables directly in terms of a small set of PNS parameters. In this vein, quantities such as luminosities and mean energies are written as functions of six global time-independent free PNS parameters, the ALP parameters, and the entropy profile. The entropy itself is obtained by solving the first-order differential equation that governs the energy-balance condition of the cooling PNS [see Eq.~\eqref{eq:difeq}], after which all observables can be expressed analytically in terms of the PNS and ALP inputs.

The six protoneutron star global parameters are the mass $M_{\rm PNS}$, radius $R_{\rm PNS}$, density correction factor $g$, opacity boosting factor $\beta$, total neutrino emitted energy $E_{\rm tot}$, and proton fraction $Y_p$. The new-physics sector includes the axion couplings and, in this work, we additionally incorporate the dependence on the axion mass. In our analysis we focus on the nucleon-nucleon Bremsstrahlung channel, for which the relevant ALP parameters are the axion-nucleon coupling $g_{aNN}$ and the ALP mass $m_a$.

Let us now summarize the key ingredients and expressions of the original setup before turning to the modifications induced by a finite axion mass. In this framework, local PNS profiles such as the temperature and density can be expressed in terms of the six global PNS parameters. For instance, the temperature profile, which depends on the entropy evolution (encoding the time dependence) and on the radial coordinate, can be written as~\cite{Suwa:2020nee}
\begin{eqnarray} \label{eq:temp}
T &=& 30 \, \MeV \, \qty(\frac{M_{\rm PNS}}{1.4 M_{\odot}})^{2/3} \qty(\frac{R_{\rm PNS}}{10 {\rm km}})^{-2}   \nonumber\\
&& \times \qty(\frac{s}{1 k_{B} {\rm baryon}^{-1}}) \qty(\frac{\sin \xi }{\xi})^{2/3} \,,
\end{eqnarray}
where the variable $s$ is the entropy per nucleon (with $k_B$ the Boltzmann constant) and we defined the dimensionless radial coordinate
\be
\xi \equiv \frac{r}{R_{\rm PNS}} \pi \, ,
\ee 
with $r$ the physical radius. Throughout this work, stellar masses are expressed in units of the solar mass $M_\odot$, and the normalizations are defined with respect to representative PNS benchmark values.

The density profile follows from the $n=1$ Lane--Emden equation of state~\cite{Weinberg:1972kfs,Suwa:2020nee}, yielding
\be \label{eq:rho}
\rho = \frac{M_{\rm PNS}}{4 \pi^2}  \qty(\frac{\pi}{R_{\rm PNS}})^3 \frac{\sin \xi}{\xi}\, . \\[2pt]
\ee

The one-flavor neutrino luminosity is approximated by the scaling relation~\cite{Suwa:2020nee}
\begin{eqnarray} \label{eq:nulum}
L_\nu &\approx& 1.2 \times 10^{50} \, {\rm erg} \, {\rm s}^{-1} \, \left(\frac{M_{\rm PNS}}{1.4 M_{\odot}}\right)^{4/5}
\left(\frac{R_{\rm PNS}}{10\, {\rm km}}\right)^{-6/5} 
\nonumber\\
&&\times  \left(\frac{g \beta}{3}\right)^{-4/5}  \left(\frac{s}{1\, k_{B} \, {\rm baryon}^{-1}}\right)^{4/5}    \,,
\end{eqnarray}
where $g$ parametrizes deviations of the PNS equation of state from the Lane–Emden case with polytropic index $n=1$ and the boosting factor $\beta$ accounts for coherent scattering of neutrinos on heavy nuclei, which enhances the effective neutrino cross-section.

Now, considering only SM processes, the PNS thermal energy~\cite{Suwa:2020nee},
\begin{eqnarray}
E_{\rm th} &=& 2.5 \times 10^{52}  \, {\rm erg} \, \qty(\frac{M_{\rm PNS}}{1.4 M_{\odot}})^{5/3} \qty(\frac{R_{\rm PNS}}{10 {\rm km}})^{-2}
\nonumber \\
&&\times \qty(\frac{s}{1 k_{B} {\rm baryon}^{-1}})^{2} \,,
\end{eqnarray}
is released predominantly in the form of neutrinos during the explosion, allowing one to relate the energy loss to the neutrino luminosity. However, in the presence of additional light degrees of freedom, such as axion-like particles, the energy budget is shared, and the loss equation becomes
\be \label{eq:difeq}
\frac{d E_{\rm th} }{ d t} = - 6 L_\nu - L_a \,,
\ee
where the factor of $6$ accounts for the three neutrino flavors and their antiparticles, and $L_a$ represents the total ALP luminosity. In what follows we focus on axion emission via nucleon–nucleon Bremsstrahlung, $L_a \simeq L_a^{\rm brem}$, and neglect subdominant channels (e.g., Primakoff~\cite{PhysRevD.37.1237}~\footnote{Note that for heavier ALP masses the photo-production channel may become relevant~\cite{Chakraborty:2024tyx}.}) unless stated otherwise.

\subsection{Including the axion mass} \label{sec:Maxion}

The coupling of axion-like particles with nucleons is described by the following Lagrangian~\cite{Carena:1988kr,Chang:1993gm,DiLuzio:2020wdo}
\be
\mathcal{L}_{aNN}= \frac{g_{aNN}}{2 m_N} (\bar \psi_N \, \gamma_\mu \gamma_5 \, \psi_N) \partial^\mu a \,,
\ee
where $g_{aNN}$ is the axion–nucleon coupling, $m_N$ is the nucleon mass, $\psi_N$ denotes the nucleon Dirac field with $N = p, n$, and $a$ is the pseudo-scalar ALP field. We assume a universal axion–nucleon coupling $g_{aNN}$ for both protons and neutrons. Since the supernova core is strongly neutron rich, the $nn$ Bremsstrahlung channel gives the dominant contribution.

In the OPE (one-pion-exchange) approximation~\cite{Brinkmann:1988vi,osti_6276659}, \textit{i.e.}, considering that the nucleons interact with each other by the exchange of one pion, the rate of energy loss per unit volume and time, $Q_a$, from nucleon--nucleon Bremsstrahlung $N\,N \to N\,N\,a$ is given by~\cite{Raffelt:1996wa,Carenza:2019pxu}
\begin{equation}
\begin{aligned}
    Q_a =
    \int & d\Pi_a \, \omega_a \int  \prod_{i=1}^{4} d\Pi_i \,
    f_1 f_2 \, (1-f_3)(1-f_4) \, (2\pi)^4  \\
    & \, \delta^{4}(p_1+p_2-p_3-p_4-p_a)\,
    S \sum_{\rm spins} |{\cal M}|^2 \, ,
\end{aligned}
\end{equation} \\ [-5pt]
where the Lorentz-invariant phase–space element is, as usual, $d\Pi_i = d^3 \bm{p}_i/[(2\pi)^3\, 2E_i]$. The four-momenta of the nucleons are written as $p_i=(E_i,\bm{p}_i)$, with
$p_{1,2}$ referring to the incoming nucleons and $p_{3,4}$ to the outgoing ones,
while $p_a=(\omega_a,\bm{p}_a)$ corresponds to the emitted axion–like particle.
The functions $f_i$ denote the phase–space occupation numbers of particle $i$,
$S$ is a symmetry factor accounting for identical particles in the initial and final
states, and the matrix element squared $|{\cal M}|^2 $ is summed over the spins of the initial and final states.

The full phase–space integral in $Q_a$ is difficult to compute, especially because of the uncertainties in nuclear interactions in the dense supernova medium. For this reason, it is customary to adopt approximations to simplify the expression. A common one is the massless–axion limit, where the ALP energy is taken as $ \omega_a \simeq |\mathbf{p}_a| $. This is a good approximation when the axion mass is much smaller than the typical temperature $T$ in the PNS. However, for ALPs in the MeV range the axion mass $m_a$ becomes relevant. In this case the dispersion relation becomes $ \omega_a = \sqrt{|\mathbf{p}_a|^2 + m_a^2} $, which changes both the available phase space and the thermal weighting of the emitted ALPs.

To proceed with the computation of $Q_a$ we adopt the following simplifications: (i) we treat nucleons as non–degenerate, so the Pauli–blocking factors $(1-f_3)(1-f_4)$ can be neglected; (ii) the phase–space distributions $f_i$ are approximated by Maxwell–Boltzmann statistics, $f_i \simeq e^{-E_i/T}$; (iii) the matrix element is computed in the OPE approximation, neglecting the pion mass. Under these approximations, and including the axion mass, we can write $Q_a$ as~\cite{Raffelt:1996wa,Lella:2022uwi}
\begin{equation} \label{eq:Qafull}
\begin{aligned}
    Q_a =  \, \frac{g_{\scaleto{aNN}{4pt}}^2}{16 \pi^2} \frac{n_B}{ m_N^2}  \int  & d \omega_a \, \omega_a \, e^{\scaleto{-\omega_a/T}{7pt}} (\omega_a^2 - m_a^2)^{\frac32} \\
    & \times \mathcal{S_\sigma} \, \scaleto{\Theta (\omega_a - m_a)}{10pt} \,,
\end{aligned}
\end{equation}
where $n_B = 2 \int f_i(\bm{p}) \, d^3\bm{p}/(2\pi)^3$ is the baryon number density, with the factor of $2$ accounting for the two spin states, $\Theta(\omega_a - m_a)$ enforces the kinematic threshold for axion emission,
and $\mathcal{S}_\sigma$ is the nucleon spin–density structure function that encodes the matrix-element information. The quantitative determination of the nucleon structure function $\mathcal{S}_\sigma$ is nontrivial, as it involves significant uncertainties from nuclear interactions and many–body effects~\cite{Carenza:2019pxu}. Since our goal is to obtain a semi-analytic expression that captures the correct physics while keeping the computation simple, we do not attempt a full numerical evaluation of $\mathcal{S}_\sigma$. Instead, we adopt the naive OPE approximation, under which the structure function can be written as
\be
\mathcal{S}_\sigma = \frac{\Gamma_\sigma\, {\hat s}(\omega_a/T)}{\omega_a^{2}} \,,
\ee
where $\Gamma_\sigma$ is the nucleon spin–fluctuation rate induced by nuclear
collisions and $\hat s$ is a dimensionless function of the ratio $\omega_a/T$.
In the non-degenerate limit these quantities are given by~\cite{Raffelt:1996wa}
\be
\begin{aligned}
& \Gamma_\sigma = 4 \, \pi^{3/2} \left( \frac{g_{\pi NN}}{m_\pi} \right)^{\!4} \rho \, T^{1/2} m_N^{1/2} \,, \\[4pt]
& \hat s(x) \simeq \sqrt{1 + \frac{\pi |x|}{4}} \,,
\end{aligned}
\ee
where $g_{\pi NN} \sim 1$ is the dimensionless pion–nucleon coupling and $m_\pi$ is the pion mass. 

The axion luminosity $L_a$ from nucleon–nucleon Bremsstrahlung is obtained by integrating the local energy–loss rate over the PNS volume,
\be
L_a = \int_0^{R_{\rm PNS}} (4 \pi r^2 ) \, Q_a \, dr \, . \\[3pt]
\ee
Using the approximations introduced above and rewriting the phase–space integral in terms of the dimensionless variable $x= \omega_a/T$, the luminosity can be expressed as
\be \label{eq:Lafull}
\begin{aligned}
L_a = &  \, \frac{g_{\scaleto{aNN}{4pt}}^2}{4 \pi^{7/2} \, m_N^{5/2}} \qty(\frac{g_{\pi NN} }{m_\pi})^4  \int_0^{R_{\rm PNS}} dr \, (4 \pi r^2 ) \, \rho^2 \, T^{\, 7/2}   \\
& \times \frac{1}{f_{\rm sup}} \, \int_{m_a/T}^\infty  \, dx \, e^{-x}   \, x^2  \,  \qty(1 - \frac{m_a^2}{T^2 \, x^2})^{3/2} \, ,
\end{aligned}
\ee
where we introduced the suppression factor $f_{\rm sup}$ to account for effects beyond the OPE approximation. As shown in Ref.~\cite{Carenza:2019pxu}, including, for example, a finite pion mass in the propagator, $\rho$-meson exchange, medium modifications of the nucleon mass, and multiple scattering effects leads to a reduction of the axion luminosity. Their Fig.~15 provides the comparison between the axion luminosity computed in the OPE approximation and in their full analysis, from which we extract the suppression factor $f_{\rm sup}$. In our final plots we allow this factor to vary in order to illustrate its impact on the resulting bounds.

Let us comment that, since the temperature also depends on the radius [see Eq.~\eqref{eq:temp}], the luminosity must be computed numerically by evaluating Eq.~\eqref{eq:Lafull}. Nonetheless, in the limit of vanishing ALP mass, the integral over the variable $x$ becomes analytic and yields a value of $2$, while the radial integral can also be performed exactly by using the analytic profiles for temperature and density given in Eqs.~\eqref{eq:temp} and~\eqref{eq:rho}. In this limit, we obtain the following analytic expression for the luminosity in terms of the PNS variables,
\be  \label{eq:Lama0}
\begin{aligned}
\lim_{m_a \to 0} L_a =& \, g_{aNN}^2 \, \times 10^{72} \, {\rm erg} \, {\rm s}^{-1} \, \qty(\frac{M_{\rm PNS}}{1.4 M_{\odot}})^{13/3}  \\
& \times \qty(\frac{R_{\rm PNS}}{10 {\rm km}})^{-10} \qty(\frac{s}{1 k_{B} {\rm baryon}^{-1}})^{7/2}    \,,
\end{aligned}
\ee
where we used $m_\pi = 135 \MeV$ and $m_N = 938 \MeV$, and $g_{aNN} = 1$.

\begin{figure}[t]
\includegraphics[width=8.5cm]{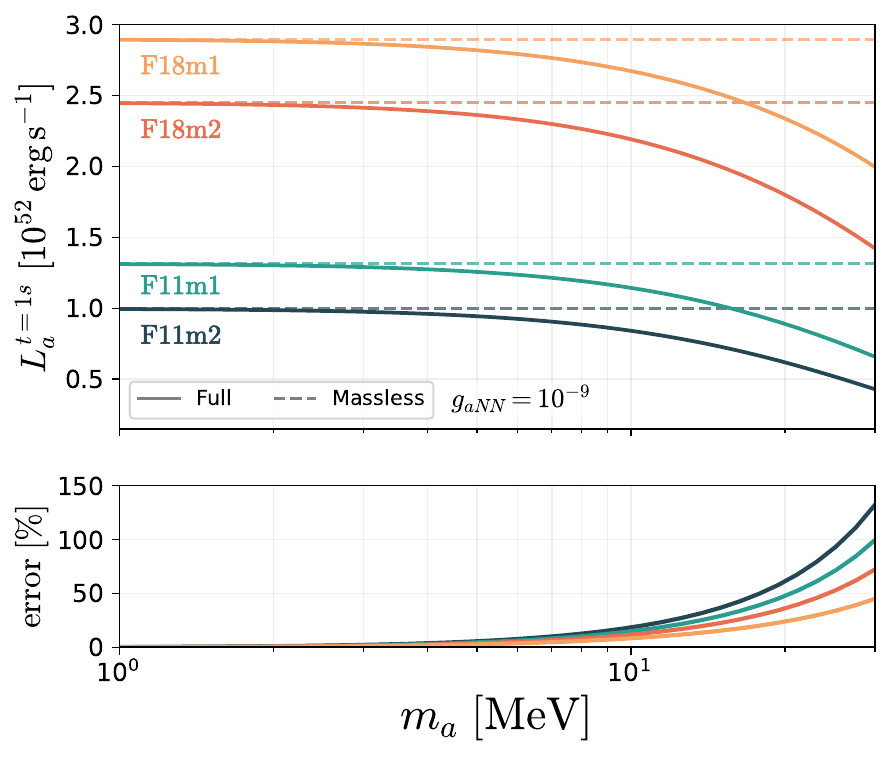}
\caption{\label{fig:Lmacomp} Upper panel: ALP nucleon-nucleon Bremsstrahlung luminosity, evaluated at $t = 1~\mathrm{s}$ post-bounce, as a function of ALP mass $m_a$. The luminosity is computed using the full numerical expression in Eq.~\eqref{eq:Lafull} (solid curves) and in the massless ALP limit given in Eq.~\eqref{eq:Lama0} (dashed curves). The different colors correspond to different choices of the PNS global parameters, as indicated in the labels and summarized in Table~\ref{tab:Fmodels}. The axion-nucleon coupling is fixed to $g_{aNN} = 10^{-9}$. Lower panel: percentage deviation obtained when the massless approximation is used instead of the full computation, for the different PNS models, following the same color scheme.}
\end{figure}

To quantify the impact of the ALP mass on the luminosity, the upper panel of Fig.~\ref{fig:Lmacomp} shows the full luminosity at $t = 1~\mathrm{s}$ post–bounce, computed from Eq.~\eqref{eq:Lafull} (solid lines), compared with the massless analytic limit of Eq.~\eqref{eq:Lama0} (dashed lines), as a function of the ALP mass. For illustration, the ALP–nucleon coupling is fixed to $g_{aNN} = 10^{-9}$. Results are shown for different choices of PNS parameters corresponding to the Fischer11 and Fischer18 models, each evaluated for two masses (see Table~\ref{tab:Fmodels}), with colors indicating the specific model. The lower panel displays the percentage error introduced when using the massless limit instead of the full computation. One can see that, for ALP masses above $\sim 10~\mathrm{MeV}$, the massless approximation overestimates the luminosity by about $10\%$ or more, independently of the model.

\subsection{Calibration with supernova simulations}\label{sec:SNsim}

As previously described, the axion luminosity in Eq.~\eqref{eq:Lafull} depends on the temperature and density profiles given in Eqs.~\eqref{eq:temp} and~\eqref{eq:rho}, which in turn are determined by the PNS global parameters together with the entropy profile. The entropy is obtained by solving the differential energy-loss Eq.~\eqref{eq:difeq} in conjunction with the neutrino luminosity expression in Eq.~\eqref{eq:nulum}, which also depends on the PNS parameters. Therefore, in order to compute the luminosity for a given ALP mass and coupling, one must specify the six global PNS parameters ($M_{\rm PNS}$, $R_{\rm PNS}$, $g$, $\beta$, $E_{\rm tot}$, $Y_p$). 

Let us note also that, in practice, the opacity-boosting factor $\beta$ and the total neutrino-emitted energy $E_{\rm tot}$ are treated separately for the early and late phases of the PNS evolution. This distinction is needed because, at early times, heavy nuclei have not yet formed in the outer layers of the PNS~\cite{Suwa:2013mva}, implying that the coherent-scattering enhancement of the neutrino opacity does not operate. As a result, the boosting factor $\beta$ acquires an explicit time dependence.

The eight PNS parameters (accounting for the early- and late-time $\beta$ and $E_{\rm tot}$) are determined by fitting the neutrino luminosity curves obtained from core-collapse supernova simulations available in the literature~\footnote{The fit is performed using the minimum $\chi^2$ method with the \texttt{Minuit} minimizer from the \texttt{iminuit} Python package~\cite{iminuit}.}. In this work we focus on two specific models, Fischer18 and Fischer11, corresponding to simulations of $18\,M_\odot$ and $11.2\,M_\odot$ pre-collapse progenitors, respectively~\cite{Fischer:2016cyd}. Since the full parameter set contains eight quantities, we allow only a subset of them to vary in the fit, while fixing the remaining ones to the values provided by the simulations. In particular, we take the PNS radius $R_{\rm PNS}$, the initial $E_{\rm tot}^i$ and final $E_{\rm tot}^f$ neutrino-emitted energy, and the late-time opacity-boosting factor $\beta_f$ to be free parameters, while the others are held fixed. The early-time boosting factor is fixed to $\beta_i = 3$ for all models, reflecting the absence of coherent scattering enhancement before heavy 
nuclei form in the PNS. Table~\ref{tab:Fmodels} summarizes the PNS parameter values extracted from the fits to the neutrino luminosity curves of the Fischer simulation models. Each model is fitted for two fixed PNS masses, chosen to match the baryonic or gravitational mass of the PNS at the end of the corresponding simulation, as reported in Table~II of Ref.~\cite{Fischer:2016cyd}.

\begin{table*}[th!]
\centering
\def\arraystretch{1.5}
\begin{center}
    \begin{tabular}{c c c c c c c c c}

         $\quad$ Model $\quad$ & $ \, \myr{M_{\rm PNS}} \, [M_\odot] \, $ & $ \quad \myg{R_{\rm PNS}} \, [{\rm km}] \quad $ & $\quad \myr{g} \quad $ & $ \quad \myr{\beta_i} \quad $ &$ \quad \myg{E_{\rm tot}^i} \, \rm{[erg]} \quad $ & $ \quad \myg{\beta_f} \quad $ &$ \quad \myg{E_{\rm tot}^f} \, \rm{[erg]} \quad $ & $ \quad \myr{Y_p} \quad $ \\
        \hline 
         \textbf{F11m1} & $1.19$ & $23.74$ & $0.04$ & $3$ & $2.82 \times 10^{52} \, $ & $66.49$ & $ 8.62 \times10^{52}$ & 0.3\\
         \textbf{F11m2} & $1.29$ & $25.75$ & $0.04$ & $3$ & $2.82 \times 10^{52} \, $ & $66.57$ & $ 8.62 \times10^{52}$ & 0.3\\
         \textbf{F18m1} & $1.46$ & $22.06$ & $0.04$ & $3$ & $3.65 \times 10^{52}$ & $70.58$ & $ 11.76 \times 10^{52}$ & 0.3\\
         \textbf{F18m2} & $1.62$ & $24.51$ & $0.04$ & $3$ & $3.65 \times 10^{52}$ & $70.69$ & $ 11.76 \times 10^{52}$ & 0.3\\
    \end{tabular}
    \caption{PNS global parameters obtained from the fit to the neutrino luminosity curves of the numerical Fischer11 (F11) and Fischer18 (F18) simulations~\cite{Fischer:2016cyd}, each evaluated at two different fixed PNS masses. Parameters shown in red (green) were kept fixed (allowed to vary) during the fit.}
    \label{tab:Fmodels}
\end{center}    
\vspace{-10pt}
\end{table*}

\begin{figure}[t]
\includegraphics[width=0.43\textwidth]{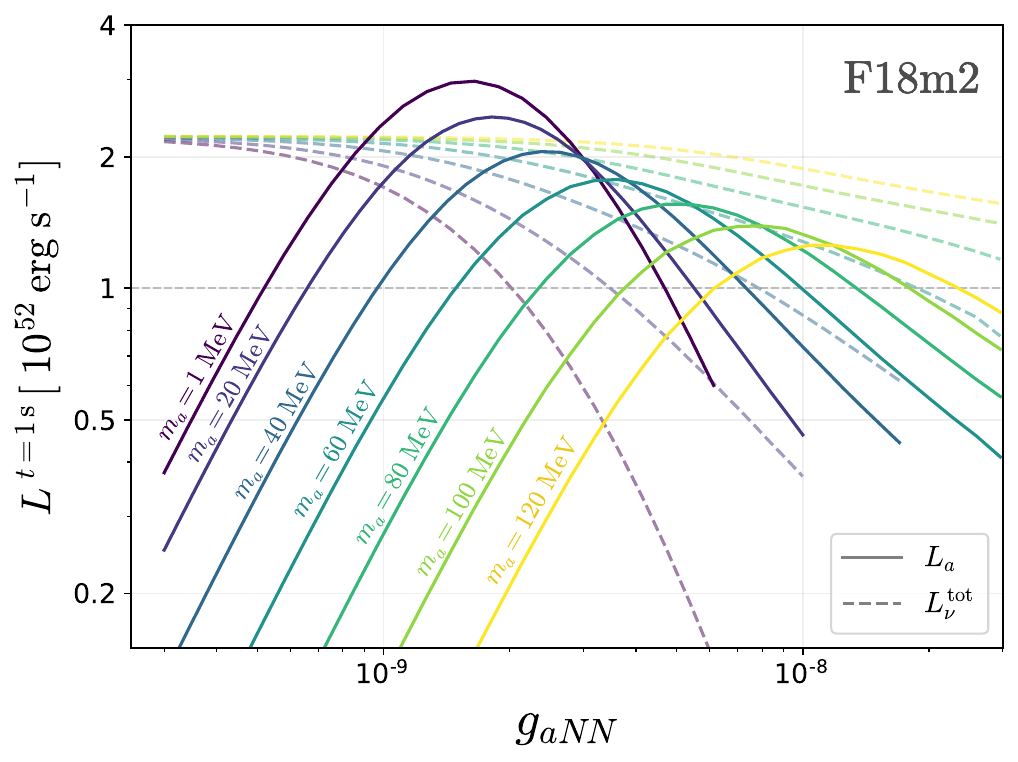}
\includegraphics[width=0.43\textwidth]{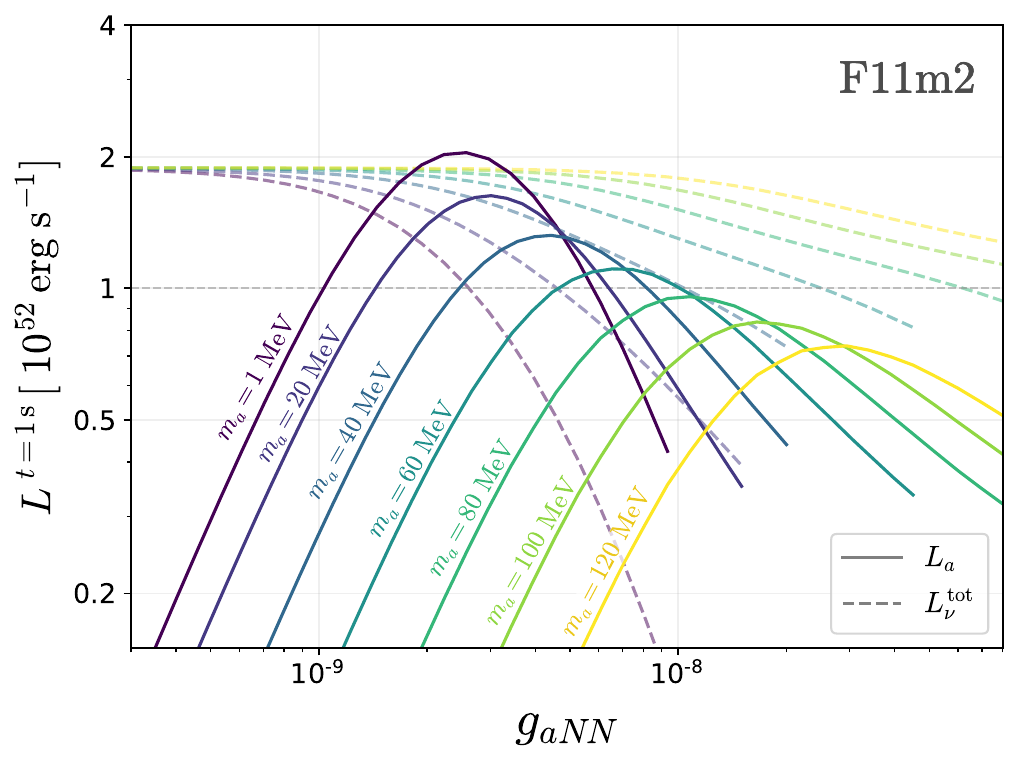}
\caption{\label{fig:LalpLnu} ALP (solid) and neutrino (dashed) luminosities at $t = 1~\mathrm{s}$ post-bounce as functions of the axion-nucleon coupling for the F18m2 model (top panel) and F11m2 model (bottom panel). The colored curves correspond to different axion masses in the range $1$--$120~\mathrm{MeV}$, as indicated by the labels.}
\end{figure}

Figure~\ref{fig:LalpLnu} shows the ALP (solid) and neutrino (dashed) luminosities, evaluated at $t = 1~\mathrm{s}$ post-bounce, as functions of the axion-nucleon coupling $g_{aNN}$ for the F18m2 (top panel) and F11m2 (bottom panel) models. The colors represent different choices of ALP masses, as indicated by the labels. One can see that for very small couplings the ALP luminosity is suppressed, as $L_a \propto g_{aNN}^2$ in the free-streaming regime. A similar suppression appears again at large couplings. This qualitative behavior resembles the well-known turnover expected when ALPs become trapped in the PNS~\cite{PhysRevD.42.3297,Lella:2023bfb}. However, in the present work we do not implement a full trapping calculation based on the axion mean free path and optical depth. Instead, the decrease of $L_a$ at large $g_{aNN}$ arises from the backreaction of axion emission on the PNS thermodynamics. Increasing the coupling enhances the local emissivity, which accelerates cooling through the energy-balance equation. This leads to a faster reduction of the entropy and therefore of the temperature at the evaluation time. Since the emissivity depends sensitively on temperature, the decrease of $T$ at large couplings eventually overcomes the $g_{aNN}^2$ scaling and causes $L_a$ to decline. 

Given that our goal here is to keep the semi-analytic framework as simple as possible, we do not include the trapping regime explicitly. Our results will therefore focus on the lower branch of the exclusion curve in the ALP coupling-mass plane, as discussed in the next section.

\section{Results}\label{sec:results}

After calibrating the global PNS parameters through fits to the reference neutrino luminosity profiles of the two Fischer progenitor simulations, we solve the differential energy-loss Eq.~\eqref{eq:difeq} for fixed ALP mass and coupling in order to determine the entropy evolution. Using the resulting entropy profile, we then compute the neutrino and ALP luminosities semi-analytically as functions of time for fixed ALP parameters. By scanning over a range of ALP masses and couplings, we are therefore able to derive bounds by imposing suitable supernova-cooling criteria.

We consider two distinct supernova-cooling criteria inspired by the original Raffelt condition~\cite{PhysRevLett.60.1793,Raffelt:2006cw}:

\begin{enumerate}[label=(\roman*)]
\item The maximum ALP luminosity $L_a(t_{\rm max})$, reached at time $t = t_{\rm max}$, is required to be smaller than the total neutrino luminosity at the same time,
\begin{equation}
L_a(t_{\rm max}) < L_\nu(t_{\rm max}) \, .
\end{equation}

\item The ALP luminosity evaluated at $t = 1~\mathrm{s}$ post--bounce is required to be smaller than the total neutrino luminosity at $t = 1~\mathrm{s}$,
\begin{equation}
L_a(1~\mathrm{s}) < L_\nu(1~\mathrm{s}) \, .
\end{equation}
\end{enumerate}

\begin{figure}[t]
\includegraphics[width=\columnwidth]{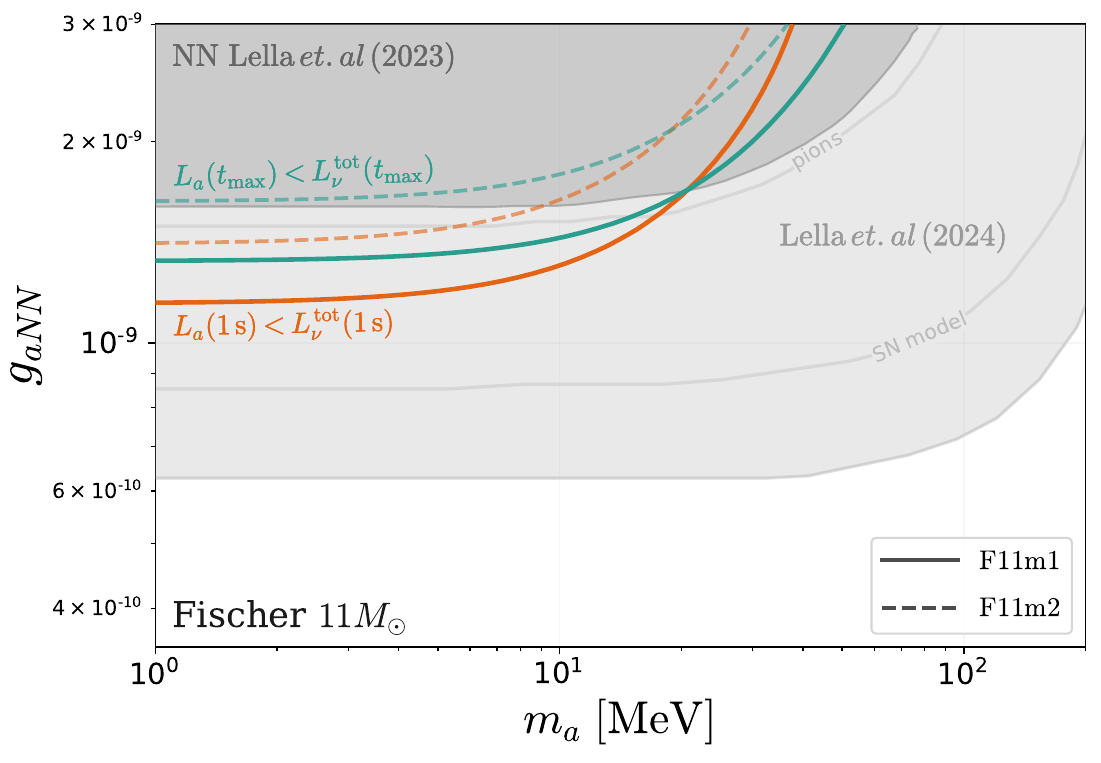}
\caption{\label{fig:F11plot} Supernova cooling bounds in the axion-nucleon coupling $g_{aNN}$ versus mass $m_a$ parameter space obtained using the fitted PNS global parameters from models \textbf{F11m1} (solid) and \textbf{F11m2} (dashed), and for the different cooling criteria (i) (green) and (ii) (orange); see the text and Table~\ref{tab:Fmodels} for details. We compare our results with the reference bounds from Ref.~\cite{Lella:2022uwi} (darker shaded region) and Ref.~\cite{Lella:2023bfb} (lighter shaded region). For the latter, we also quote their results when adopting a different SN model simulation (grey solid ``SN model'' curve) and when including pion contributions in the SN core (grey solid ``pions'' curve). }
\end{figure}

Figure~\ref{fig:F11plot} shows the bounds in the $(g_{aNN}, m_a)$ plane obtained using criteria (i) (green) and (ii) (orange) for the models \textbf{F11m1} (solid) and \textbf{F11m2} (dashed). For comparison, we also display previous bounds from the literature. The darker shaded region corresponds to the nucleon-nucleon Bremsstrahlung bound (labeled NN Lella \textit{et al.} (2023)) shown in Fig.~3 of Ref.~\cite{Lella:2022uwi}. The lighter shaded region corresponds to the updated bound presented in Fig.~3 of Ref.~\cite{Lella:2023bfb} (Lella \textit{et al.} (2024)). 

In addition, we include the curves shown in Fig.~4 of Ref.~\cite{Lella:2023bfb} to illustrate the sensitivity of the exclusion region to different modeling choices. One curve reflects the impact of adopting an alternative supernova simulation (AGILE--BOLTZTRAN instead of the reference Garching model) and is labeled ``SN model'' in Fig.~\ref{fig:F11plot}. The other curve shows the modification obtained when pion contributions in the SN core are included and is labeled ``pions''. These comparisons highlight the level of systematic uncertainty associated with the supernova modeling and the underlying microphysics.

From Fig.~\ref{fig:F11plot} one can see that the bounds obtained using our framework overlap significantly with the regions previously excluded in the literature. In particular, our exclusion curves lie within the band defined by the reference results of Refs.~\cite{Lella:2022uwi,Lella:2023bfb}, indicating good agreement despite the different methodological approaches. Let us emphasize again that our bounds are obtained via a semi-analytic framework without the need of computationally expensive simulations. Therefore, the consistency of our results with the previous bounds in the literature indicates the robustness of the method developed here.

\begin{figure}[t]
\includegraphics[width=\columnwidth]{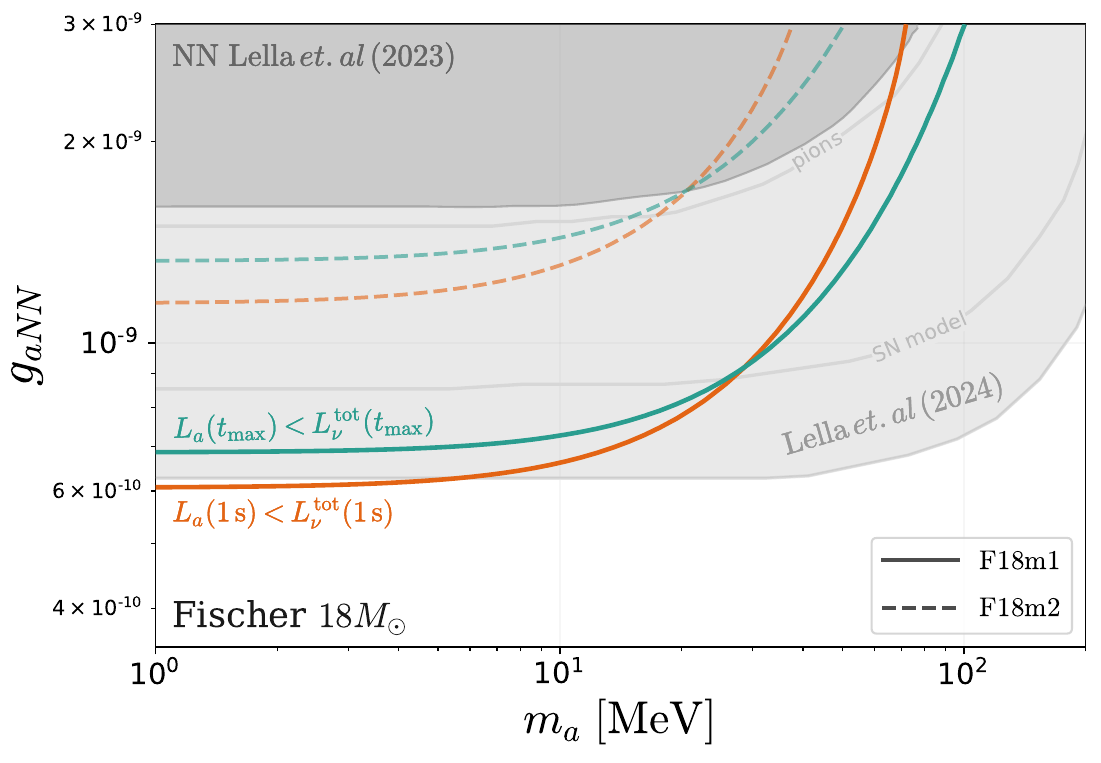}
\caption{\label{fig:F18plot}
Same as Fig.~\ref{fig:F11plot}, but using the fitted PNS global parameters from the Fischer18 simulations: \textbf{F18m1} (solid green and orange) and \textbf{F18m2} (dashed green and orange). The remaining curves are identical to those shown in Fig.~\ref{fig:F11plot}.
}
\vspace{3mm}
\end{figure}

Figure~\ref{fig:F18plot} shows the same as Fig.~\ref{fig:F11plot}, but for the bounds obtained using the Fischer18 simulations, i.e., employing the PNS global parameters extracted from the \textbf{F18m1} (solid colored) and \textbf{F18m2} (dashed colored) models. The green and orange curves correspond again to cooling criteria (i) and (ii), respectively, and the reference comparison bounds are the same as in Fig.~\ref{fig:F11plot}. One can see that the Fischer18 model, corresponding to a larger progenitor mass, yields more constraining bounds compared to the Fischer11 case. This is expected, since a heavier progenitor leads to a more massive and hotter PNS, which enhances the ALP emissivity and therefore tightens the exclusion. Nevertheless, the obtained exclusion regions remain in good agreement with the reference limits reported in the literature.

\begin{figure}[t!]
\includegraphics[width=\columnwidth]{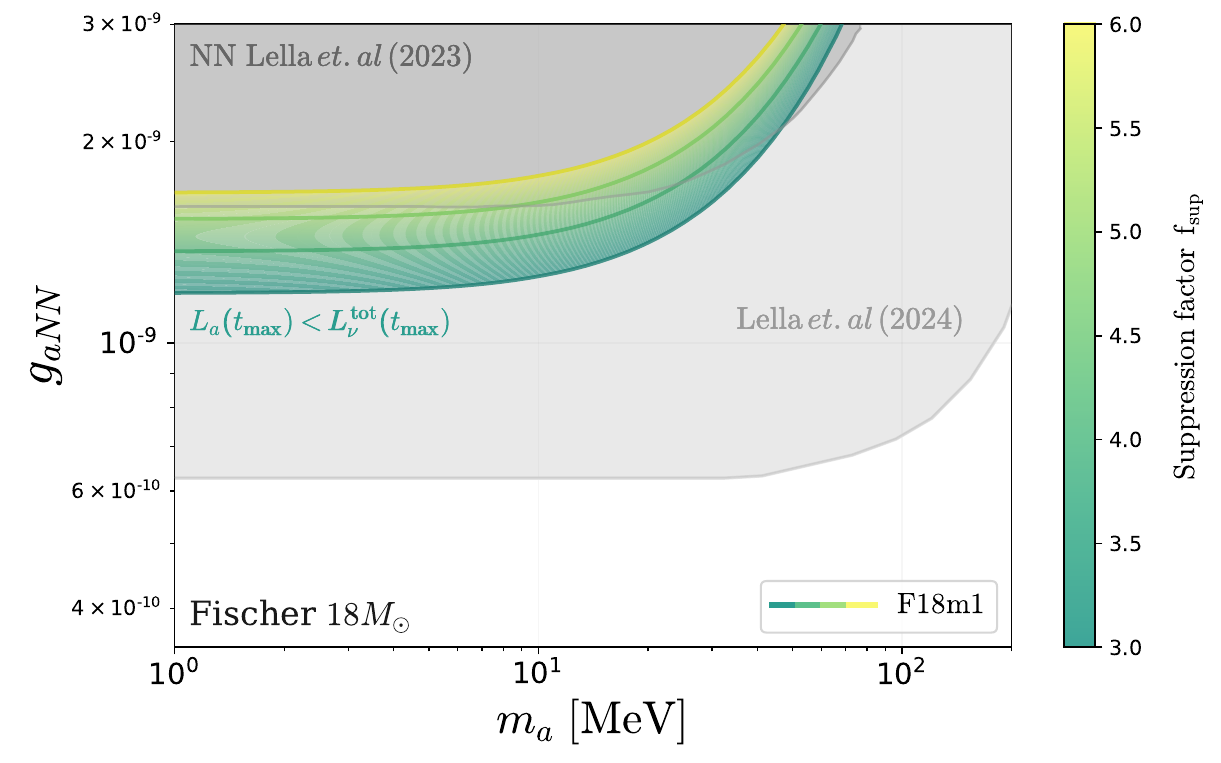}
\caption{\label{fig:Fsupplot}
Supernova-cooling bounds in the $(g_{aNN}, m_a)$ plane obtained using the \textbf{F18m1} model and cooling criterion (i), while varying the suppression factor in the range $f_{\rm sup} \in [3,6]$. The color gradient indicates the value of $f_{\rm sup}$, and the highlighted curves correspond to the representative values $f_{\rm sup}=3,4,5,6$. The background exclusion regions correspond to the same reference bounds from the literature shown in Fig.~\ref{fig:F11plot}.
}
\end{figure}

Finally, in Fig.~\ref{fig:Fsupplot} we illustrate the impact of including the suppression factor $f_{\rm sup}$ in the ALP luminosity computation, as shown in Eq.~\eqref{eq:Lafull} and discussed in Sec.~\ref{sec:Maxion}. The colored curves correspond to bounds obtained with the \textbf{F18m1} model using cooling criterion (i), while varying the suppression factor in the range $f_{\rm sup}=3$ (darker green) to $f_{\rm sup}=6$ (yellow). The color bar indicates the value of $f_{\rm sup}$ across the shaded contour region, and the highlighted curves correspond to the representative values $f_{\rm sup}=3,4,5,6$.~\footnote{Let us note that in Ref.~\cite{Carenza:2019pxu}, in the weak-coupling regime, the OPE approximation overestimates the axion luminosity by approximately a factor of six, corresponding to a representative suppression factor $f_{\rm sup} \sim 6$. In our analysis we allow for variations around this value in order to reflect residual nuclear uncertainties. } The background exclusion regions are the reference bounds from the literature, shown for comparison as in Fig.~\ref{fig:F11plot}. From the figure one can see that, as expected, increasing the suppression factor shifts the exclusion curve toward larger values of the coupling. Physically, a larger $f_{\rm sup}$ reduces the axion luminosity for fixed $(g_{aNN}, m_a)$, so a stronger coupling is required to satisfy the same cooling criterion. As a result, the bound becomes weaker.

\vspace{-5pt}
\section{Outlook} \label{sec:outlook}

The extreme conditions reached in core-collapse supernovae make these transient events powerful laboratories for probing new physics. However, the high complexity of the involved microphysics makes full numerical simulations of such events highly demanding and computationally expensive. Besides, the significant astrophysical and nuclear uncertainties, combined with the lack of observational data, imply that the derivation of precise constraints on new-physics models from supernova cooling arguments remains subject to considerable systematic uncertainties, even when state-of-the-art simulations are employed.

In this context, semi-analytic methods offer a practical and computationally inexpensive alternative for estimating new-physics observables and constraints. By employing the framework developed in Ref.~\cite{Foguel:2022fef,Suwa:2020nee}, we extended the semi-analytic description of ALP emission in core-collapse supernovae to incorporate a finite axion mass, and used it to derive exclusion bounds in the axion-nucleon coupling versus mass plane, $(g_{aNN},m_a)$.

The semi-analytic framework expresses all relevant PNS observables, such as the temperature, neutrino luminosity, and thermal energy, as functions of six global time-independent PNS parameters and the entropy, which carries the full time dependence of the system. The entropy evolution is then obtained by solving the energy-loss differential equation for a given set of ALP parameters. The global PNS parameters were calibrated by fitting the semi-analytic neutrino luminosity curves to those extracted from state-of-the-art core-collapse supernova simulations. Here we considered two different progenitor model simulations with $18\,M_\odot$ (Fischer18) and $11.2\,M_\odot$ (Fischer11) pre-collapse masses.

The exclusion bounds in the $(g_{aNN},m_a)$ plane obtained with this approach are in good agreement with previously published results~\cite{Lella:2022uwi,Lella:2023bfb} derived from full numerical simulations, demonstrating the robustness and utility of the semi-analytic method. We compared bounds obtained across the two progenitor calibrations and two distinct cooling criteria, finding consistent results in all cases. We also investigated the sensitivity of the bounds to nuclear effects beyond the one-pion-exchange approximation, which suppress the axion luminosity relative to the OPE result. These effects were parametrized by the suppression factor $f_{\rm sup}$, such that by varying it we quantified the sensitivity of the bounds to these residual nuclear uncertainties in the axion emissivity.

Let us note that in this work we did not include the trapping regime, which becomes relevant at large axion-nucleon couplings and would allow one to derive the upper branch of the exclusion contour. Its inclusion would require a dedicated calculation, especially since at such large couplings the presence of axions could significantly affect the supernova explosion mechanism itself~\cite{Betranhandy:2022bvr}, potentially compromising the validity of the standard cooling argument. The assessment of this region is therefore left to future work. In addition, recent studies suggest that future measurements could constrain part of that upper region~\cite{Alonso-Gonzalez:2024spi}. Therefore, the focus of the present work was on the lower branch of the exclusion curve, corresponding to the free-streaming regime.

To summarize, we showed that the semi-analytic approach developed here provides a simple and fast alternative for deriving bounds and testing new physics, compared to computationally expensive full numerical simulations. Given the significant uncertainties in the astrophysical and nuclear microphysics, the fact that our semi-analytic bounds fall within the regions previously obtained in the literature demonstrates the robustness of the method. Furthermore, the flexibility of the semi-analytic framework makes it well-suited for extensions to other ALP production channels and different new physics scenarios, which will be explored in future work.

\vspace{-5pt}
\begin{acknowledgments}
We would like to thank Alessandro Lella and Giuseppe Lucente for useful discussions. This work was partially supported by INCT-FNA (Process No. 464898/2014-5), CAPES (Finance Code 001), CNPq, and FAPERJ. ALF is supported by Funda\c{c}\~ao de Amparo \`a Pesquisa do Estado de S\~ao Paulo (FAPESP) under the contracts 2022/04263-5, and 2024/06544-7.
\end{acknowledgments}

\bibliography{SNnu}

\end{document}